\documentclass[journal=aamick,manuscript=article, layout=twocolumn]{achemso}
\usepackage{chemformula}
\usepackage[T1]{fontenc}

\author{Sara Memarzadeh}
\email{sarmem@amu.edu.pl}
\affiliation{Institute of Spintronics and Quantum Information, Faculty of Physics and Astronomy, Adam Mickiewicz University Poznań, Uniwersytetu Poznańskiego 2, 61-614 Poznań, Poland}
 
\author{Mateusz Gołębiewski}
\affiliation{Institute of Spintronics and Quantum Information, Faculty of Physics and Astronomy, Adam Mickiewicz University Poznań, Uniwersytetu Poznańskiego 2, 61-614 Poznań, Poland}

\author{Maciej Krawczyk}
\affiliation{Institute of Spintronics and Quantum Information, Faculty of Physics and Astronomy, Adam Mickiewicz University Poznań, Uniwersytetu Poznańskiego 2, 61-614 Poznań, Poland}

\author{\\Jarosław W. Kłos}
\affiliation{Institute of Spintronics and Quantum Information, Faculty of Physics and Astronomy, Adam Mickiewicz University Poznań, Uniwersytetu Poznańskiego 2, 61-614 Poznań, Poland}

\title[An \textsf{achemso} demo]{Nucleation and Arrangement\\ of Abrikosov Vortices\\ in Hybrid Superconductor-Ferromagnetic Nanostructure}


\begin{document}

\begin{abstract}
This study investigates the nucleation, dynamics, and stationary configurations of Abrikosov vortices in hybrid superconductor-ferromagnetic nanostructures exposed to inhomogeneous magnetic fields generated by a ferromagnetic nanodot. Using time-dependent Ginzburg-Landau simulations and Maxwell’s equations, we observe the evolution of curved vortex structures that exhibit creep-like deformation 
while reaching a steady state. Spatial variations in the Lorentz force, along with the interaction between geometric constraints and vortex interactions, give rise to unusual stationary vortex configurations that gradually change with increasing field strength, a behavior not seen in homogeneous magnetic fields. These findings reveal complex pinning mechanisms, providing valuable insights for the optimization and further advancement of nanoscale superconducting systems. 
\end{abstract}

\section{Introduction \label{sec:Intro}}
In type-II superconductors (SCs), an external magnetic field induces a lattice of vortices containing quantized magnetic flux, known as Abrikosov vortices, when it exceeds the critical value $H_{c1}$\cite{PhysRev.108.1175, Abrikosov1957}. This effect is particularly interesting for SCs of finite dimensions, especially those with sizes on the order of a few superconducting correlation lengths\cite{Xu_2008, Baek_2015}. It has been shown that the finite size can reduce the critical field, and the shape of the SC can influence the regions of vortex nucleation and the steady-state vortex arrangement, which may differ from the triangular lattice observed in the extended SC. For example, the studies in Refs.~\citenum{Berdiyorov_2009, Peng2017, peng_2016} shed light on the complex interplay between magnetic and geometrical factors in SCs, particularly by comparing the vortex states in cubic and spherical geometries as a function of the magnetic field. In these cases, the vortices may prefer a curved shape instead of a straight, columnar shape as seen in films. In addition, the orientation of a homogeneous magnetic field with respect to the surfaces of the SC influences the vortex configurations\cite{Brandt1993}, as shown for the pyramidal SC structures~\cite{hasnat_2020}. Such studies of the spatial arrangement and dynamics of vortices are essential for understanding the response of SCs to magnetic fields\cite{Tinkham1996, kopnin2001theory} and their applications, especially in IT and quantum technologies where miniaturization is crucial\cite{DeTeresa_2023}.
However, to date, most research has focused on homogeneous magnetic fields in both 2D SC wires\cite{Schweigert_1998, Alstrøm2011} and 3D SC nanodots\cite{Peng2017, peng_2016, hasnat_2020, Aguirre_2017, gonzalez_2015, AGUIRRE20241354450}.

The combination of ferromagnetic (FM) and SC nanostructures, either by direct proximity effects\cite{RevModPhys.77.935} or in separation via electromagnetic fields\cite{Bespalov_2022}, offers new possibilities to control the configurations and dynamics of both subsystems. These include the influence of SC currents and vortices on the magnetization configuration in the FM\cite{Palau_2016,Gonzalez_2022}, or the effect of the stray magnetic field from the FM on the SC state \cite{Milosevic_2004,Jafri_2020,Dahir_2020,Niedzielski2020}. 
Additionally, the mutual coupling between FM and SC, e.g., involving chiral magnetic textures and superconductivity, 
enable for the strong interactions between the magnetic skyrmion and the SC vortex (anti-vortex)\cite{PhysRevLett.126.117205,Gonzalez_2022}.
Dynamic coupling can also be observed in SC-FM hybrids. For example, the inhomogeneous RF magnetic field of a single magnetic dipole can induce vortex semiloops in the SC nanoelement \cite{Oripov2020}, which evolve over time, or the domain wall propagating in FM layer can push the vortices in electromagnetically coupled SC layer\cite{PhysRevB.109.L201110}. On the other hand, the stray field generated by the SC nanoelement or the lattice of vortices can affect the magnetization dynamics in the FM layer\cite{Dobrovolskiy_2019,Borst_2023,Kharlan_2024}.

Despite these important contributions, there are still significant gaps in the research on SC-FM hybrids. These include those related to the physics of hybrid systems where both SC and FM are at the nanoscale, particularly involving inhomogeneous magnetic fields. The nanoscale size, comparable to the SC correlation length, and the inhomogeneous static stray field generated by the FM nanodot affect the nucleation of vortices, their transient dynamics and their stable arrangements. Theoretical studies on vortex configurations in the stationary state for planar SC systems exposed to nonuniform magnetic field reveal the complexity of such problem\cite{Menezes_2019} and make us aware that for SC-FM nano-hybrids the following issues should also be considered: (i) the 3D geometry of the SC system, which goes beyond the simplicity of planar structures, (ii) the finite and small sizes of the SC structure, within the range of a few correlation lengths, where the external shape plays a significant role in the whole volume of the SC, (iii) the pinning mechanisms resulting from the complex energy landscape in the confined geometry while stabilizing the vortex configuration, and (iv) the inhomogeneous stray magnetic field from the finite size FM, which adds additional complexity to the SC system.

This paper fills these gaps by a comprehensive numerical study of the influence of spatially inhomogeneous magnetic fields on the vortex nucleation, their transient dynamics, and stable arrangement. Specifically, we investigate 3D SC nanodots exposed to the inhomogeneous magnetic fields generated by nearby permanent nanomagnet, with both having the same lateral dimensions. We solve the time-dependent Ginzburg-Landau (TDGL) equations \cite{Ginzburg2009, gulian2020shortcut} and Maxwell's equations using the finite element method (FEM). We show that the inhomogeneity of the stray magnetic field significantly affects the vortex nucleation and the vortex arrangement in the SC.
Interestingly, the vortex dynamics that reach a steady state are governed by the magnetic field lines, resulting in a mixture of curved and straight vortices with the normal phase regions. These properties depend on the SC-FM separation, the lateral and vertical dimensions of the SC in terms of coherence length and penetration depth, making them important factors influencing the potential usefulness of SC-FM hybrid structures in spintronic and quantum technology applications\cite{Melnikov2022,Makarov2022,DeTeresa_2023,Cai2023}.  

The structure of the paper is as follows: in Sec.~\ref{sec:Model}, we present the theoretical model and the computational framework used for our simulations. In Sec.~\ref{sec:Res}, we discuss the results, first in subsection~\ref{sec:dynamics}, focusing on vortex nucleation, dynamics, and stationary configurations under inhomogeneous magnetic fields. Furthermore, we analyze the static configuration of the relaxed system as a function of the external field, particularly its magnetic properties (Sec.~\ref{sec:static}). By comparing these hybrid SC-FM systems with both 2D SC systems and 3D SC nanostructures under homogeneous magnetic fields, we aim to elucidate the interplay between geometry and external fields, providing new perspectives on the vortex dynamics in these hybrid systems. Finally, in Sec.~\ref{sec:Conc}, we summarize the conclusions drawn from this study and suggest possible directions for future research. In addition, in Supporting Information we extend the manuscript with the details of TDGL equation derivation, details of numerical modeling, and numerical results for SC vortex dynamics and arrangements in different geometries

\section{Model \label{sec:Model}}

\subsection{Method}
The dynamic and static properties of the SC system are described by the TDGL equations. The TDGL theory\cite{Gorkov, schmid} is formally restrictive to the gapless SCs. However, even for materials with a non-zero superconducting gap, the presence of impurities can broaden the singularities of density of states  and close the gap \cite{lee_2023}, restoring the validity of TDGL theory  
Therefore, the formulation of the model based on TDGL equations is widely accepted, and effective compromise between relative simplicity and accuracy in numerical studies of SC systems and devices\cite{Horn_2023} 

We solve TDGL equations with the FEM in COMSOL Multiphysics\textsuperscript{\tiny\textregistered} \cite{zimmerman2006multiphysics, comsol2009}. To simplify the simulations, we make time $t$, the complex order parameter $\psi$, and the electrical conductivity $\sigma$ dimensionless\cite{Alstrøm2011}:
\begin{equation}
    t \rightarrow \frac{\xi^2}{D} t, \quad \psi \rightarrow \sqrt{\frac{\alpha}{\beta}} \psi, \quad \sigma \rightarrow \frac{1}{\mu_0 D \kappa^2} \sigma, \label{eq:units}
\end{equation}
where $\xi$ is the coherence length, $D$ is the electron diffusion coefficient, $\alpha$ and $\beta$ are material-specific parameters, and $\mu_0$ is the permeability of free space. The GL parameter $\kappa = \lambda / \xi$ is the ratio of the London penetration depth $\lambda$ to the coherence length $\xi$. However, we do not use dimensionless units for the vector potential $\mathbf{A}$, the external magnetic field $\mathbf{B}_{\rm a}$, and the spatial coordinates $x, y, z$ to ensure proper coupling of the GL equations with Maxwell's equations, which describe the stray field produced by the FM element.

With these transformations, the TDGL equation for order parameter ($\psi$) becomes:
\begin{align}
    \frac{\partial \psi}{\partial t}
    = - \frac{\lambda^2}{\kappa^2}(i \nabla+\frac{q}{\hbar}\mathbf{A})^{2} \psi + \psi - \lvert \psi \rvert^{2} \psi,\label{eq:psi}
\end{align}
and the equation for the vector potential ($\mathbf{A}$) is given by:
\begin{align}
    \sigma \frac{\partial \mathbf{A}}{\partial t}
    = & \frac{\hbar}{2iq} (\psi^{*}\nabla \psi-
    \psi \nabla \psi^{*})
    -\lvert {\psi} \rvert ^{2} A 
    \nonumber \\ &
    - \lambda \nabla \times ( \lambda \nabla \times \mathbf{A}- \mathbf{B}_{a}).\label{eq:vecpot}
\end{align}
Here, $q = 2e$ is the charge of the Cooper pair, and $\hbar$ is the reduced Planck's constant. For further details, please refer to the Supporting Information, Secs.~S1 and S2.

For simplicity, we assume that the nanomagnet, with high saturation magnetization and large out-of-plane uniaxial anisotropy, is uniformly magnetized and the relatively small field produced by SC prism does not affect the static magnetization of the nanomagnet (see Supporting Information Sec.~S3, in particular, Fig.~S2). The magnetic field generated by nanomagnet $\mathbf{B}_{\text{FM}}$ is calculated within the magnetostatic approximation, considering the magnetization within the FM prism oriented along the $z$-axis:
\begin{equation}
    \mathbf{B}_{\rm FM}=-\mu_0\nabla \varphi_{M}, \quad \Delta \varphi_{M}=\nabla\cdot\mathbf{M},
\end{equation}
where $\varphi_{M}$ is a magnetostatic potential (additional information is available in Supporting Information, Sec.~S2). 

The vortex configuration within the SC prism, in particular, when exposed to the nanomagnet field, is expected to be complex, often containing poorly formed, and weakly isolated vortex structures that may merge with regions of the normal phase. Therefore, the number of vortices or even the vorticity, is no longer a reliable parameter for characterizing the SC nanoelement\cite{Berthod_2005}. Instead, we quantitatively evaluate the screening properties of these SC nanostructures by their diamagnetic response, as reflected in the volume- or surface-averaged magnetization\cite{Berdiyorov_2009, nino_2019, gonzalez_2015, Aguirre_2017, hasnat_2020}. This represents the difference between the total magnetic field [$\nabla \times \mathbf{A}(\mathbf{r})$] and the applied field [$\mathbf{B}_{\rm a}(\mathbf{r})$] within the SC and serves as an indicator of the demagnetizing properties: 
\begin{equation}
    \langle\left|\mathbf{M}\right|\rangle = \frac{1}{\mu_{0}}\frac{1}{ V_{\text{SC}} } \int_{V_{\text{SC}}} \left| \nabla \times \mathbf{A}(\mathbf{r}) - \mathbf{B}_{\rm a}(\mathbf{r}) \right|d^3\mathbf{r},\label{eq:magnetisation}
\end{equation}
where $V_{\text{\rm SC}}=a\times a\times h$ is the volume of the SC prism. In two-dimensional simulations, the volume integrals are adapted to the surface integrals.

To study vortex formation and arrangement, we define regions where the supercurrent, $|\psi|^2$, is significantly reduced, which will indicate vortices and normal phase indentations. It is clear that there is no sharp boundary between these two phases, but we arbitrarily introduce a threshold value, i.e., $|\psi|^2 = 0.3$, for effective identification of boundaries between the SC and normal phases, as illustrated in Fig.~S3 in the Supporting Information. To quantify this, we introduce a filling fraction $f\!f_N$, which is the ratio of the volume in which the density of Cooper pairs $|\psi|^2$ is less than 0.3 to the volume of the SC prism:
\begin{equation}
    f\!f_N = \frac{1}{V_{\text{SC}}} \int_{V_{\text{SC}}} \Theta\left(0.3 - |\psi(\mathbf{r})|^2\right) d^3\mathbf{r}. \label{eq:ff}
\end{equation}
Here, $\Theta(x)$ represents the Heaviside step function. In other words, the parameter $f\!f_N$ represents the volume in which superconductivity is reduced (i.e. the volume of vortices and indentations of normal phase), relative to the total volume of the SC prism.

In FEM simulations, we use a high-quality discretization mesh with triangular elements in 2D and tetrahedral elements in 3D. The linear size of the mesh elements ranged from $\varepsilon_{\rm min} = 0.25~\text{nm}$ to $\varepsilon_{\rm max} = 25$~nm. More details about the mesh and the derivation of TDGL equations (\ref{eq:psi}) and (\ref{eq:vecpot}), as well as the details concerning the numerical procedure, like a convergence test and boundary conditions, are presented in the Supporting Information, Sec.~S1 and S2.

\subsection{Structure}
\begin{figure}[!htbp]
  \centering
    \begin{tabular}{c}
      \includegraphics[width=\linewidth]{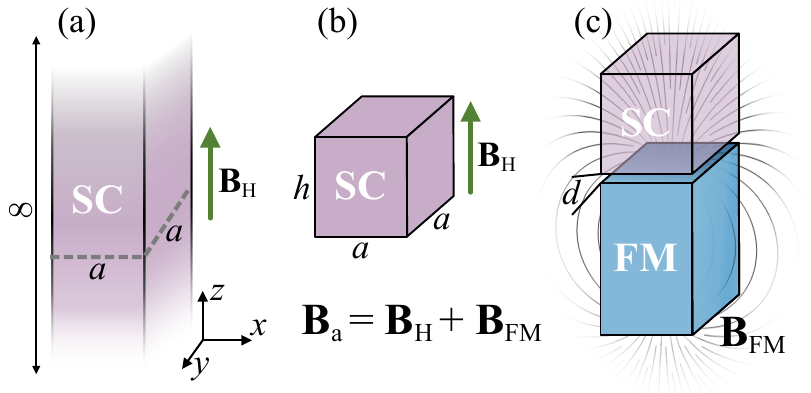}
    \end{tabular}
\caption{Numerical studies investigate vortex configurations in a SC prism under an external magnetic field $\mathbf{B}_{\rm a}$. The reference systems are: (a) an infinite SC wire and (b) a prism of height $h$ in a homogeneous magnetic field $\mathbf{B}_{\rm H}$. (c) We focus on the effect of the inhomogeneous field $\mathbf{B}_{\rm FM}$, generated by the FM prism, on the SC prism. The SC and FM prisms are separated by an air gap of width $d$, and throughout all studies, both the SC and FM components have the same cross-sectional area of $a \times a$.}
\label{fig:structure}
\end{figure}

We focus on the hybrid SC-FM structure shown in Fig.~\ref{fig:structure}(c) where both the SC and FM components have a cross-sectional area of $a \times a$. The size $a$ and the height $h$ of the SC prism are varied, while the height of the FM prism (nanomagnet) is fixed at $h_M = 700$~nm. The SC is characterized by a GL parameter $\kappa = 3$, a penetration depth $\lambda = 60$~nm, and a dimensionless electrical conductivity, $\sigma=1$ [expressed in the units $1/(\mu_0 D \kappa^2)$]. To obtain a relatively strong magnetic field from the FM, $\mathbf{B}_{\text{FM}}$, we selected its saturation magnetization to be $M_{\rm s} = 1350$~kA/m. We tune the strength of the inhomogeneous magnetic field in the SC by adjusting the distance $d$ between the SC and the nanomagnet. As its measure, we use the average stray magnetic field of the nanomagnet over the SC volume: $\langle |\mathbf{B}_{\rm a}| \rangle = \frac{1}{V_{\text{SC}}} \int_{V_{\rm SC}} B_{\rm FM}(\mathbf{r}) d^3\mathbf{r}$.

In simulations, we use the time scale in $\tau$ units, defined in Eq.~\ref{eq:units}, $\tau=\xi^2/D$. With a penetration depth $\lambda = 60$~nm and a GL parameter $\kappa = 3$, the coherence length becomes $\xi = \frac{\lambda}{\kappa} = 20$~nm. Using literature values for $D$, we estimate $\tau$ for several SCs\cite{halbritter1971,meservey1971,cohen1968}. For aluminum, with a diffusion coefficient of $1 \times 10^{-4} \, \text{m}^2/\text{s}$, the characteristic time scale is $\tau_{\text{Al}} = 4$~ps; for niobium, with $D = 1 \times 10^{-6} \, \text{m}^2/\text{s}$, $\tau_{\text{Nb}} = 400$~ps, and for lead, with $D = 5 \times 10^{-7} \, \text{m}^2/\text{s}$, $\tau_{\text{Pb}} = 800$~ps. These time scales indicate how quickly the phenomena in the SC evolve, with dynamics occurring on the order of picoseconds for aluminum and on the order of hundreds of picoseconds for niobium and lead.

To understand the vortex nucleation and the stable configuration in the SC-FM hybrid, we consider three additional reference systems. The first one is an infinite SC wire of the square cross-section under a homogeneous magnetic field $B_{\rm H}$ [Fig.~\ref{fig:structure}(a)] with the dimensions $a \times a$, the same as the selected SC-FM hybrid structure. This system was modeled using a 2D model confined to the $xy$-plane. The infinite extension of this geometry along the $z$ direction ensures that the SC system does not generate a stray magnetic field outside the boundary of the SC. It not only provides a baseline for a better understanding of vortex behavior in more complex scenarios but also serves to validate the accuracy of our computational model\cite{Alstrøm2011, gulian2020shortcut}. 
The second reference system is an SC prism under a homogeneous magnetic field $B_{\rm H}$ [Fig.~\ref{fig:structure}(b)], already studied in the literature\cite{Peng2017}, with the same dimensions as the studied SC-FM system. Here, a 3D model was used to capture the already complex vortex arrangement. The final reference system is an SC sphere in a homogeneous magnetic field, which allows the differences in vortex nucleation and their stable arrangement, determined by the SC shape, to be shown. 

\section{\label{sec:Res}Results and Discussion}

The study is divided into two parts. In the first part, we study the process of vortex nucleation and its time evolution to a stable configuration in SC prism placed in nonuniform field of nanomagnet [Fig.~\ref{fig:structure}(c)]. In the second part, we focus on the stable vortex configuration as a function of the distance $d$ between the SC prism and the FM nanodot, i.e., as a function of the mean magnitude of the magnetic field $\langle |\mathbf{B}_{\rm FM}| \rangle$ or of $\mathbf{B}_{\text{H}}$. The results concerning the stable configuration are compared to the outcomes for the reference systems [Fig.~\ref{fig:structure}(a)], [Fig.~\ref{fig:structure}(b)] and the SC sphere (Sec.~S6 in the Supporting Information).

\subsection{\label{sec:dynamics} Vortex nucleation and transient dynamics }
\begin{figure*}[!htbp]
    \begin{tabular}{c c}
         \includegraphics[page=2,width=.45\textwidth]{pics/time_evolution.pdf}& 
         \makebox[0.32\textwidth][l]{
           \hspace{3mm}
            \raisebox{0.5cm}[0pt][0pt]{  
         \includegraphics[width=0.3\linewidth, height=0.4\linewidth]{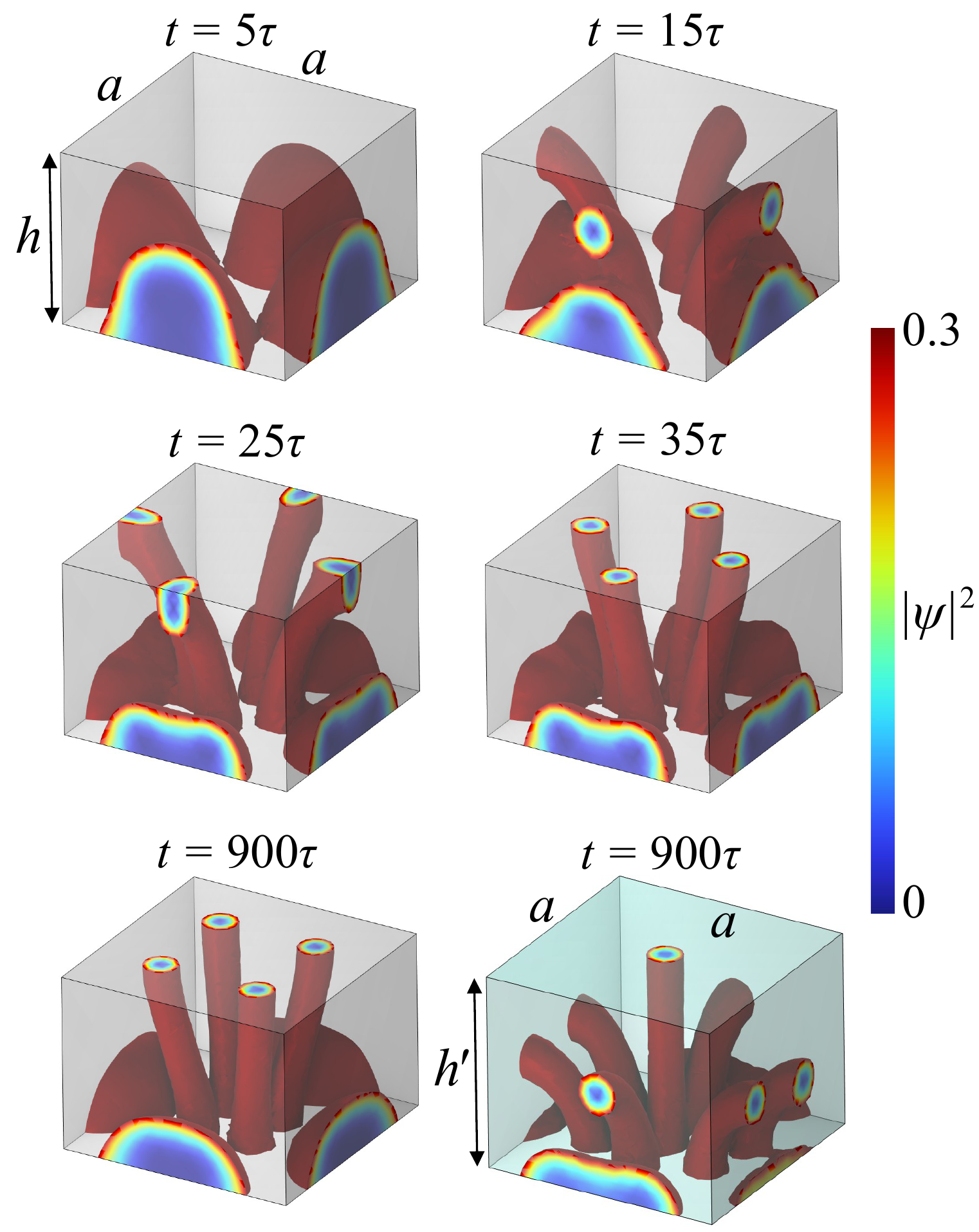}  
         }}
    \end{tabular}
    \caption{
Evolution of the SC prism under homogeneous and inhomogeneous magnetic fields. The upper left graph shows the time-dependent behavior of the filling fraction $f\!f_{N}$. The lower left graph illustrates the time evolution of the system's magnetization $\langle |\mathbf{M}| \rangle$. The right-hand panels depict the evolution of 
the flux region and vortex structure in a 3D SC over time, under an inhomogeneous magnetic field from a nearby ferromagnetic nanodot at a distance of $d = 10$~nm. The final snapshots at $t=900\tau$, with heights $h = 185$~nm and $h' = 205$~nm, show the static system's vortex structures. The SC prism has a square cross-section with side length $a = 250$~nm.
The SC-FM distance remains constant at $d = 10$~nm for both structures, leading to magnetic field strengths of $\langle |\mathbf{B}_{\rm FM}| \rangle = 336$~mT for $h = 185$~nm and $ 315 $~mT for $h' = 205$~nm. For the SC prism in a homogeneous magnetic field $\mathbf{B}_{\rm H}=|\mathbf{B}_{\rm FM}|$.
}
    \label{fig:vortex_dynamics}
\end{figure*}

The dynamics of the system are assessed during the transition from an initial state, in which the prism is entirely in a SC state, characterized by $\psi(t=0) = 1$ throughout its volume, to stable configuration, where $\tfrac{d}{dt}\psi\rightarrow 0$. At time $t=0$, an external magnetic field, $\mathbf{B}_{\rm a}$ (produced either by ferromagnetic nanodot $\mathbf{B}_{\text{FM}}$ or uniform one $\mathbf{B}_{\text{H}}$), is suddenly applied, prompting a magnetic response of the SC material. The system is then allowed to evolve freely until equilibrium is reached, where variations in the order parameter $\psi$ and the vector potential $A$ are sufficiently small to be considered constant over time. 

The results of such simulations are summarized in Fig.~\ref{fig:vortex_dynamics} (left part) showing $f\!f_{N}$ and $\langle |\mathbf{M}| \rangle$ as a function of time for the SC prism in the inhomogeneous field [Fig.~\ref{fig:structure}(c)], drawn with red lines, and for the reference for the SC prism in the uniform field [Fig.~\ref{fig:structure}(b)], drawn with blue lines. The vortex configurations (represented as regions where $|\psi|^2 < 0.3$) are shown at selected times on the right part of Fig.~\ref{fig:vortex_dynamics}. The prism has fixed dimensions: $250 \times 250 \times 185$ nm$^3$ and the mean magnetic field is 336~mT. The results for the higher prisms, i.e., $250 \times 250 \times 205$ nm$^3$, are shown for comparison in Supporting Information, Fig.~S5. 

Based on the $f\!f_{N}(t)$ dependence, we can notice that the transition in the SC-FM structure goes through four distinct stages before reaching a steady state, while in the SC prism under a homogeneous magnetic field, there are only three distinct stages. In the SC-FM hybrid system, the process starts from the Meissner phase throughout the whole SC, similar to the reference systems. The magnetic flux begins to penetrate the SC prism at the midpoints of the bottom edges forming the indentations with a low $|\psi|^2$ (see the right part of the Fig.~\ref{fig:vortex_dynamics} where the regions $|\psi|^2<0.3$ are marked at $t=5 \tau$), which increase with time. This process is quantified by the increase of $f\!f_{N}$ (the upper left graph in Fig.~\ref{fig:vortex_dynamics}). Before the transition to the mixed state, the system reaches its maximum $f\!f_{N} \cong 0.25$ at $t=5 \tau$. 

In the next stage, vortices emerge from the nucleation points localized at the tops of the indentations of the normal phase (see Supporting Information, Fig.~S4 at $t=10$ $\tau$), gradually forming curved vortices that elongate and change their curvature as their ends climb up the lateral faces of the SC prism (see the vortex configuration at $t=15\tau$). Eventually, all the vortices reach the top surface of the prism at $t=25 \tau$. The analysis of the $f\!f_{N}(t)$ dependence reveals that this parameter decreases significantly until the end of this stage $t=25 \tau$.
The dynamics in the stage, when the curved vortices are nucleated and grow towards the top face of the SC prism, i.e., from $t = 5\tau$ to $25\tau$ can be considered, by analogy with mechanical\cite{Huda2022}, ferromagnetic\cite{Lemerle_1998} or ferroelectric\cite{Tybell_2002} systems, as a creep-like deformation\cite{RevModPhys.66.1125} (see also the discussion in Sec.~S7 of the Supporting Information). 

These two initial stages are significantly different for the SC prism exposed to a homogeneous external magnetic field, i.e., the structure shown in Fig.~\ref{fig:structure}(b) (see the red curves in Fig.~\ref{fig:vortex_dynamics}). Here, the columnar vortices nucleate quite abruptly from the sides of the prism (at $t \approx 5\tau$) and are fully formed with the lengths determined by the height of the structure. In this case, the time from the nucleation to the formation of column vortices is only about $\Delta t=5 \tau$ compared to the much longer time $\Delta t=15 \tau$ of creep-like dynamics in the inhomogeneous case (see plot of $f\!f_{N}(t)$ in Fig.~\ref{fig:vortex_dynamics}). Thus, as expected, in the system with a homogeneous magnetic field, straight vortices are energetically favorable (see also Fig.~S4 in the Supporting Information) \cite{Beloglazov_1992}. 

In the third stage of the SC-FM heterostructures evolution towards a stable vortex configuration,
the vortices migrate inwards from the edges of the top surface during the time change from $t=25 \tau$ to $t=35 \tau$ (see Fig.~\ref{fig:vortex_dynamics}). As a result, the initially curved vortices gradually straighten out but remain curved even in the final phase (see the visualization at $t=900 \tau$). In this stage, we observe a further reduction of $f\!f_{N}$ to $\cong 0.17$, If the prism height is increased while keeping the lateral dimensions the same, more vortices remain strongly curved and their outlets do not reach the top face of the prism in the stationary state as shown in the lower right sub-figure of Fig.~\ref{fig:vortex_dynamics} and also in Fig.~S5 of the Supporting Information.

In the final stage of the evolution, just before reaching the static state, the vortex columns begin to rearrange, rotating by 45~deg around the vertical axis of the prism (compare the plots of $|\psi|^2$ in Fig.~\ref{fig:vortex_dynamics} at $t=35\tau$ and $t=900 \tau$). In this phase, the filling factor decreases only slightly to the value $f\!f_{N}\cong 0.16$. A similar rotation and rearrangement of the columnar vortices is also observed in the SC prism in uniform field, as shown in Fig.~S4 in Supporting Information, Sec.~S4. However, here the $f\!f_{N}$ remains almost unchanged and at a much lower level, i.e., $f\!f_{N}\cong 0.08$. This behavior optimizes vortex penetration, minimizes repulsive vortex energy, and also creates space for vortex nucleation when the bias magnetic field is increased (see discussion in the next section).

The above-described changes in $f\!f_{N}$ during nucleation and vortex stabilization are followed by changes in the magnetization $\langle |\mathbf{M}| \rangle$, as shown in the lower left plot in Fig.~\ref{fig:vortex_dynamics}. For the SC prism in a homogeneous magnetic field, the magnetization of SC prism sharply declines from the high value $\sim\!213$~kA/m to 18~kA/m at $\Delta t=8 \tau$, following vortex formation and stabilizes before the vortices undergo rotation. In contrast, for the SC prism placed in an inhomogeneous magnetic field (SC-FM hybrid structure), the decrease in magnetization is more gradual, from $\sim\!86$~kA/m to 18~kA/m over $\Delta t=26 \tau$. However, unlike the $f\!f_{N}$ curve, the time evolution of the magnetization does not show distinct intermediate steps.

Behind the described vortex nucleation and their formation in a columnar structure, there is a complex mechanism driven by the interplay between  applied magnetic field, SC currents, vortex interactions, and the geometry and size of the SC body\cite{Buchacek2019, Nelson1995, Baert_1995}. Nevertheless, the observed differences between the SC prism in non-uniform and uniform external magnetic fields [see the direct comparison of the vortex configuration relaxation under both homogeneous and inhomogeneous external fields in Supporting Information, Sec.~S4] suggest a correlation between the vortex bending and the direction of the external magnetic field lines. Although we cannot provide a complete description and explanation of this effect, we propose intuitive arguments below to help understand it.

In the GL free energy functional outlined in Eq.~(S1), the term $\left| \left(\nabla - \frac{iq}{\hbar} \mathbf{A}(\mathbf{r}, t) \right) \psi (\mathbf{r}, t)\right|^2$ represents the kinetic energy, which captures the interaction between the spatial variation of the order parameter and the vector potential. The last term, $\left| \mathbf{B} - \mathbf{B}_a \right|^2$, quantifies the magnetic energy density and measures the energy cost associated with the deviation of the total magnetic field $[\mathbf{B}(\mathbf{r}, t)=\nabla \times \mathbf{A}(\mathbf{r}, t)]$ from the external field $(\mathbf{B}_a)$ and it is crucial for understanding vortex stability.

When the order parameter gradient, $\nabla \psi (\mathbf{r}, t)$, associated with the SC current circulation around the vortex core, aligns with $\mathbf{A} (\mathbf{r}, t)$, it indicates that the vortex core is oriented along the local magnetic field lines $[\nabla \times \mathbf{A}(\mathbf{r}, t)]$. 
This alignment minimizes the term $i \mathbf{A}(\mathbf{r}, t) \cdot \psi ^* (\mathbf{r}, t) \nabla \psi (\mathbf{r}, t)$ in the kinetic energy, as well as $\left| \nabla \times \mathbf{A}(\mathbf{r}, t) - \mathbf{B}_a \right|^2$ in the magnetic energy, thereby increasing the stability of the vortex. Physically, if the vortices are misaligned with the applied magnetic field lines, the interaction between the magnetic field and the supercurrent associated with the vortices leads to an increase in the magnetic energy. Consequently, the vortices exhibit a strong tendency to align with the magnetic field lines to minimize energy.

However, as the vortex bends, the order parameter gradient $\nabla \psi$ undergoes rapid spatial variations, leading to an increase in $|\nabla \psi|^2$ and hence higher kinetic energy. Furthermore, in the context of the TDGL equation, the term $\nabla^2 \psi$ can be interpreted as a curvature contribution to the kinetic energy, further amplifying the energy associated with the vortices. It is important to recognize that the curvature of the vortices distorts the supercurrent distribution [as expressed in Eq.~(S5)], which is fundamentally related to $\nabla \psi$. Therefore, the presence of curved structures imposes an increase in energy, while straight configurations are energetically favored due to their ability to maintain a more uniform supercurrent distribution.

In light of this, we identified two opposing effects, one favoring the alignment of vortex lines according to the biased magnetic field lines and the other favoring straight columnar vortices. While vortices attempt to align with magnetic flux lines, columnar flux line configurations serve as the preferred structure, clarifying the tendency of vortices to migrate towards straighter flux lines during the third phase of the vortex transition process (Fig.~\ref{fig:vortex_dynamics} at $t=35 \tau$). Nevertheless, we must keep in mind that in an SC system characterized by nanoscale size and subjected to an inhomogeneous magnetic field, the vortex configuration is additionally imposed by other effects, e.g. (i) boundary conditions\cite{Brandt1993}, (ii) the inhomogeneity of the Lorentz force acting on the vortices, resulting from variations in the Meissner current density circulating in the SC material, and (iii) the vortex-vortex interactions, which depend on the absolute position of the vortices. Consequently, the stationary state of the vortex configuration is determined by the balance between the different effects, which includes both the deformation of individual vortices and their interactions \cite{Brandt1995, Kogan_1995, Gurevich_2013}.


\subsection{\label{sec:static}Magnetization in the stationary state}

\begin{figure*}
    \centering
    \begin{tabular}
    {p{0.45\textwidth} p{0.01\textwidth} p{0.45\textwidth}}
        {\includegraphics[page=1,width=.45\textwidth]{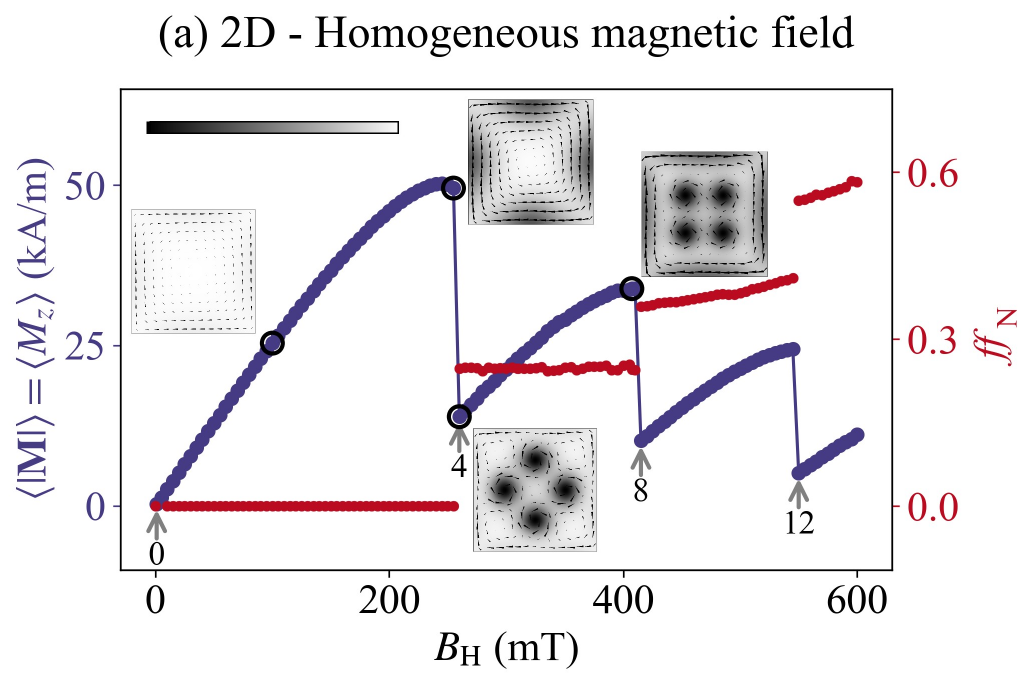}}
        \put(-142,112){{{1}}}
        \put(-200,112){{{0}}}
        \put(-175,110){{\textbf{$\left|\psi\right|^2$}}}
        & & 
         {\includegraphics[page=2,width=.45\textwidth]{pics/250_denser_denser_mesh.pdf}}
        \\ 
        \multicolumn{3}{c}{
        \includegraphics[page=11,width=.45\textwidth]{pics/250_denser_denser_mesh.pdf}
        } \\
        \includegraphics[page=3,width=.45\textwidth]{pics/250_denser_denser_mesh.pdf}
        & & 
        \makebox[0.45\textwidth][l]{ 
           \hspace{0mm}
            \raisebox{3.2cm}[0pt][0pt]{  
            \begin{tabular}
            {p{0.11\textwidth} p{0.11\textwidth} p{0.11\textwidth}}
                 \includegraphics[page=10,width=.11\textwidth]{pics/250_denser_denser_mesh.pdf}
                 \put(-60,51){\large{{i}}}&
                 \includegraphics[page=9,width=.11\textwidth]{pics/250_denser_denser_mesh.pdf}
                 \put(-60,51){\large{{ii}}}& 
                 \includegraphics[page=8,width=.11\textwidth]{pics/250_denser_denser_mesh.pdf}
                 \put(-60,51){\large{{iii}}}\\ 
                \includegraphics[page=7,width=.11\textwidth]{pics/250_denser_denser_mesh.pdf}
                \put(-60,51){\large{{iv}}}& \includegraphics[page=6,width=.11\textwidth]{pics/250_denser_denser_mesh.pdf}
                \put(-60,51){\large{{v}}}& \includegraphics[page=5,width=.11\textwidth]{pics/250_denser_denser_mesh.pdf}
                \put(-60,51){\large{{vi}}}\\ 
                \includegraphics[page=4,width=.11\textwidth]{pics/250_denser_denser_mesh.pdf}
                \put(-60,51){\large{{vii}}}& 
                \multicolumn{2}{c}{
                \raisebox{1cm}{ 
                \includegraphics[page=12,width=.20\textwidth,height=.008\textheight]{pics/250_denser_denser_mesh.pdf}} 
                \put(-105,18){{{0}}}
                \put(-8,18){{{0.3}}}
                \put(-55,16){{\textbf{$\left|\psi\right|^2$}}}
                }
                \\ 
            \end{tabular}
        }} \\ 
    \end{tabular}
\caption{Dependence of average magnetization, filling fraction, and vortex structure in SCs with different geometries on magnetic field: (a) Average magnetization $\langle |\mathbf{M}| \rangle$ (blue) and vortex density $f\!f_{N}$ (red) vs. homogeneous magnetic field $B_{\rm a}=B_{\rm H}$ for long wire. Transitions are marked by gray arrows, highlighting where the number of vortices changes. Insets show vortex patterns and supercurrent distribution as cones at different fields. (b) As in (a), but for 3D SC with homogeneous fields. Vortex density and magnetization behavior with increasing $B_{\rm a}=B_{\rm H}$ is shown. (c) 3D SC under an inhomogeneous field from a nearby FM. The graph depicts $\langle |\mathbf{M}| \rangle$ (color from color-bar corresponding to the SC-FM distance $d$) and $f\!f_{N}$ (red) vs. average magnetic field $\langle |\mathbf{B}_{\rm FM}| \rangle$. The gray dashed line represents the point at which the first vortex appears. Visualizations (i–vii) display static vortex configurations showing vortex adaptation as distance decreases. In all cases, the SC is a $250 \times 250 \, \text{nm}^2$ square prism with a height of 320~nm. 
Values $\langle ... \rangle$ are averaged over the SC volume (3D) or surface (2D).}
\label{fig:2D-3D-first system}
\end{figure*}

In this subsection, we demonstrate how the stationary state of the hybrid SC-FM system evolves as we increase the strength of the external field. The initial conditions for each simulation are set to $\psi(t=0) = 1$, with either $\mathbf{B}_{\rm a}(t=0) = \mathbf{B}_{\text{FM}}$ or $\mathbf{B}_{\text{H}}$, assuming the abrupt application of the magnetic field. For comparison, we also simulate the SC prism and infinite SC wire, both in a homogeneous magnetic field along the $z$-axis.

We first refer to the results for the SC square wire [Fig.~\ref{fig:structure}(a)] under a homogeneous magnetic field shown in Fig.~\ref{fig:2D-3D-first system}(a). 
In the Meissner state, i.e., at fields smaller than the first critical field $B_{c1}$, where no stable vortex is present, the filling fraction $ f\!f_{N} $ remains close to zero, and the magnetization is proportional to $\mathbf{B}_{\rm H}$ due to increasing Meissner current density at the wire surfaces [see the insets in Fig.~\ref{fig:2D-3D-first system}(a) at $B_{\rm H}=100$~mT and 255~mT]. At fields exceeding $B_{c1}$, each mid-side acts as an equivalent nucleation center, allowing the simultaneous nucleation of vortices in groups of four, as indicated by the numbers 4, 8, and 12 in Fig.~\ref{fig:2D-3D-first system}(a). The change in the number of vortices significantly alters the magnetization of the system and the values of the $f\!f_{N}$. The fields corresponding to the step decrease in magnetization (due to the formation of the new vortices and screened supercurrent) and the step increase in filling fraction (to $f\!f_{N} = 0.25$, 0.36, and 0.55) coincide with the minimum field values at which the system supports 4, 8, or 12 vortices in the stationary state, i.e., $\mathbf{B}_{\rm H} = 260$, 415, and 550~mT, respectively. As the field increases between the formation of additional vortices, the Meissner currents gradually increase again until the Lorentz forces acting on the vortices are strong enough to induce their 45~deg rotation around the vertical axis (compare the insets at $B_{\rm H}=260$~mT and 407~mT).

The symmetrical process of entering of vortices in groups of four can be disturbed by many factors. For example, if the magnetic field changes semi-adiabatically, i.e., with $\psi(t=0)$, each step is taken from the stationary state of the previous step at a slightly lower field. In this case, for the field step $\Delta B_{\text{H}} = 5$~mT, only the first vortices nucleate in a group of four as shown in Supporting Information, Fig.~S6. Here, the number of simultaneously nucleated vortices depends on the symmetry of the previous vortex configuration and the interactions between the vortices. Another example is the SC wire with a defect, e.g., in the form of the notch in the center of one side of the wire. It breaks the fourfold symmetry and allows individual nucleation and stabilization of vortices with increased magnetic field (see, Supporting Information, Fig.~S6). In the following paragraphs, we will see a similar effect in the SC prism placed in homogeneous and especially inhomogeneous magnetic fields.

Indeed, in the SC prism in a homogeneous field [Fig.~\ref{fig:2D-3D-first system}(b)] we observe metastable configurations with 1, 2, 3, or 5 vortices appearing in narrow field ranges\cite{Peng2017}. Notably, the magnetization of the infinite wire (maximum at 50~kA/m) generally exceeds that of the prism (maximum at 36~kA/m), indicating larger average Meissner currents. This difference can be attributed to the absence of stray fields outside the boundary in the SC wire, which enhances the Meissner effect and preserves the fourfold symmetry.

\begin{figure*}
    \centering
    \begin{tabular}{p{0.49\textwidth} p{0.01\textwidth} p{0.4\textwidth}}
        \multicolumn{3}{c}{
            \includegraphics[page=11,width=.49\textwidth]{pics/350-denser_denser_mesh.pdf}
        } \\
        \includegraphics[page=1,width=.49\textwidth]{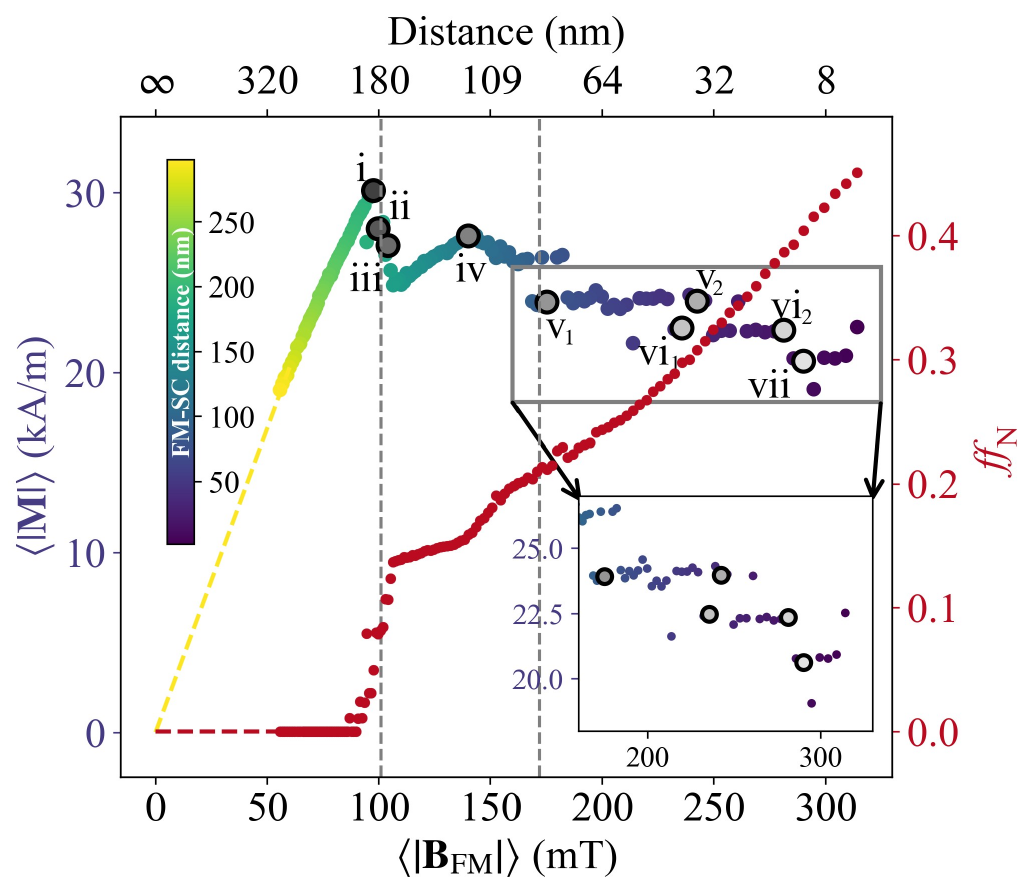} 
        & &
        \makebox[0.4\textwidth][l]{
          \hspace{-5mm}
            \raisebox{3.5cm}[0pt][0pt]{
                \begin{tabular}{p{0.11\textwidth} p{0.11\textwidth} p{0.11\textwidth}}
                    \includegraphics[page=10,width=.12\textwidth]{pics/350-denser_denser_mesh.pdf}
                     \put(-60,46){\large{{i}}}& 
                    \includegraphics[page=9,width=.12\textwidth]{pics/350-denser_denser_mesh.pdf} 
                     \put(-60,46){\large{{ii}}}& 
                    \includegraphics[page=8,width=.12\textwidth]{pics/350-denser_denser_mesh.pdf} 
                     \put(-60,46){\large{{iii}}}\\
                    \includegraphics[page=7,width=.12\textwidth]{pics/350-denser_denser_mesh.pdf} 
                     \put(-60,46){\large{{iv}}}& 
                    \includegraphics[page=6,width=.12\textwidth]{pics/350-denser_denser_mesh.pdf} 
                     \put(-60,46){\large{{$\mathrm{v_{1}}$}}}& 
                    \includegraphics[page=5,width=.12\textwidth]{pics/350-denser_denser_mesh.pdf} 
                     \put(-60,46){\large{{$\mathrm{vi_{1}}$}}}\\
                    \includegraphics[page=4,width=.12\textwidth]{pics/350-denser_denser_mesh.pdf} 
                     \put(-60,46){\large{{$\mathrm{v_{2}}$}}}& 
                    \includegraphics[page=3,width=.12\textwidth]{pics/350-denser_denser_mesh.pdf} 
                     \put(-60,46){\large{{$\mathrm{vi_{2}}$}}}& 
                    \includegraphics[page=2,width=.12\textwidth]{pics/350-denser_denser_mesh.pdf} 
                     \put(-60,46){\large{{vii}}}\\
                    \multicolumn{3}{c}{
                    \includegraphics[page=12,width=.24\textwidth]{pics/350-denser_denser_mesh.pdf}
                    }
                    \put(-162.5,-10){{{0}}}
                    \put(-49.5,-10){{{0.3}}}
                    \put(-105,-12){{\textbf{$\left|\psi\right|^2$}}}
                \end{tabular}
            }
        }
    \end{tabular}
   \caption{
   The left graph displays the average magnetization $\langle |\mathbf{M}| \rangle$ (color corresponding to the SC-FM distance) and filling factor $f\!f_{N}$ (red) plotted against the average magnetic field $\langle |\mathbf{B}_{\rm FM}| \rangle$ (SC-FM separation) for a 3D SC prism. 
   The first gray dashed line represents the first critical field, and the second gray dashed line indicates where the first curved vortex becomes straight.
   The right: visualizations (i–viii) of the static vortex configurations at selected field values.  The SC is a $350 \times 350 \text{nm}^2$ square prism and 320~nm in high.
   }
    \label{fig:3D-second system}
\end{figure*}
In the hybrid SC-FM system [Fig.~\ref{fig:2D-3D-first system}(c)], a non-uniform magnetic field introduces significant complexity in stationary vortex configuration compared to the uniform field scenario [Fig.~\ref{fig:2D-3D-first system}(b)]. This change is already evident in the rather continuous dependence of $\langle |\mathbf{M}| \rangle$ and $f\!f_N$ on the applied field strength shown in Fig.~\ref{fig:2D-3D-first system}(c). For the considered structures, $B_{c1}$ is 260~mT for the SC wire (a), 235~mT for the SC prism (b), and 100~mT for the SC-FM (c), as indicated by the left vertical dashed line in Fig.~\ref{fig:2D-3D-first system}(c). Similarly, the maximum value of $\langle |M| \rangle$ is 50~kA/m for structure (a), 36~kA/m for structure (b), and 24~kA/m for structure (c). It is evident that the magnetization values for the hybrid system decrease compared to the homogeneous field scenarios. Even the infinite wire, which theoretically provides optimal magnetic shielding, shows a response that is significantly weaker than perfect diamagnetism. This limitation is attributed to the finite London penetration depth and the small dimensions of the superconducting material, as discussed in Supporting Information, Sec.~S3.

Let us discuss in more detail the evolution of the vortex arrangement in dependence on the field of the hybrid system shown in Fig.~\ref{fig:2D-3D-first system}(c)]. For the field $\langle |B_{\rm FM}| \rangle \approx 100$~mT, we observe a stationary configuration containing four normal phase indentations located at the centers of the lower edges, where the density of Meissner currents is highest [see label (i)]. As we relax the system at slightly higher fields, we observe the emergence of one (ii) at $\langle |\mathbf{B}_{\rm FM}| \rangle = 107$, two (iii) at $\langle |\mathbf{B}_{\rm FM}| \rangle = 117$, three (iv) at $\langle |\mathbf{B}_{\rm FM}| \rangle = 125$, and finally four (v) $\langle |\mathbf{B}_{\rm FM}| \rangle =138$ curved mini-vortices, effectively expelling the remaining normal phase indentations. The selection of faces for mini-vortex formation in configurations (ii), (iii), and (iv) occurs accidentally, influenced by symmetry breaking due to numerical approximations, such as mesh asymmetry and numerical procession. In a real sample, it can be due to any structural imperfections. As more curved vortices develop, the overall screening of the SC prism diminishes, resulting in a downward trend in $\langle |M| \rangle$. As the applied field increases beyond the levels seen in configuration (v), the screening can exceed those in configurations (ii)-(v), leading to an enhanced diamagnetic response in the system. This enhancement is reflected in the increased magnetization $\langle |M| \rangle$ [see inset in Fig.~\ref{fig:2D-3D-first system}(c)]. At even higher fields, the normal phase begins to penetrate the SC prism at its bottom edges (vi), further complicating the overall screening effect.

It is important to note that the maximum field value $\langle |\mathbf{B}_{\rm FM}|\rangle \approx 250$~mT [Fig.~\ref{fig:2D-3D-first system}(c)] corresponds to a small separation of 3~nm between the nanomagnet and the SC prism. To further increase the $\langle|\mathbf{B}_{\rm FM}|\rangle$ would require unrealistically high saturation magnetization values for the nanomagnet, given the current system geometry. However, in such fields, we expect to see an expansion of the \textcolor{black}{ normal phase indentations} and/or an increase of vorticity, a deterioration of screening, and ultimately a loss of superconductivity throughout the SC sample, which could manifest in decreasing the magnetization, as observed in SC wire and SC prism.

The changes related to the emerging new vortices are also observed in $f\!f_N(\langle|\mathbf{B}_{\rm FM}|\rangle)$ dependence. Starting from configuration (i), as the system transitions into the mixed state, the filling fraction increases gradually from zero in the Meissner state. This increase signifies the penetration of the magnetic field, leading to a gradual change in SC state rather than a sudden vortex formation. These results indicate that the energy landscape in the SC-FM system is highly complex. 

We also consider how general the properties discussed in Fig.~\ref{fig:2D-3D-first system} are. Fig.~\ref{fig:3D-second system} shows results for a hybrid SC-FM system with an enlarged cross-sectional area of $350 \times 350$~nm$^2$ for both the SC prism and the nanomagnet, keeping SC height, SC-FM distance, and material parameters the same. With the increased lateral dimensions of the system, there is a greater capacity for a higher number of vortices, including those that are less curved. The enlarged cross-section of the nanomagnet produces a more uniform magnetic field at the center of the SC prism cross-section. This supports the columnar vortices, as discussed previously. Accordingly, it results in a gradual increase of the filling fraction with increasing the external field (shown as the red curve in Fig.~\ref{fig:3D-second system}), which reinforces the previous observations of continuous changes. Considering $\langle\left|\mathbf{M}\right|\rangle(B_{\rm a})$, similar behavior is observed as in the smaller SC prism, including a decrease in magnetization at the magnetic field of about 100 mT which is higher than $B_{c1}$. However, in the current system, the presence of columnar vortices induces a slight reduction in the magnetization ($\approx 2$~kA/m). The stationary vortex configurations, which contain the same number of columnar vortices, exhibit comparable magnetization as can be seen in the configurations from the groups v with one vortex (v$_1$, v$_2$), vi with two vortices (vi$_1$, vi$_2$). Overall, as the magnetic field increases above $B_{c1}$, a clear downward trend in $\langle|\mathbf{M}|\rangle$ is evident, consistent with the behavior shown in Fig.~\ref{fig:2D-3D-first system} (a) and (b).

Our results show that the energy landscape in nanoscale hybrid SC-FM systems is complex, with multiple local minima at similar magnetic fields but corresponding to different vortex configurations. It is indicated by different groups of the vortex arrangements overlapping the same range of $\langle|\mathbf{B}_{\rm FM}|\rangle$ as shown in the insets in Figs.~~\ref{fig:2D-3D-first system} and \ref{fig:3D-second system} (groups v, vi and vii). In examining this observation, we note that for a given applied field, the system can relax into several configurations defined by different numbers of columnar vortices. For instance, the configurations $({\rm v}_2, {\rm vi}_1)$ are influenced by nearby magnetic fields, suggesting that the system, starting from an initial state of $\psi=1$, can follow several equivalent paths to minimize its energy, eventually reaching different energy minima.

Furthermore, we examined the stabilization of vortices in the SC sphere of similar volume as in Fig.~\ref{fig:3D-second system} in an inhomogeneous magnetic field -- see details in the Supporting Information, Sec.~S6. We found that in dependence $\langle\left|\mathbf{M}\right|\rangle (\langle|\mathbf{B}_{\rm FM}|\rangle)$, there is a fairly clear sequence of overlapping magnetization levels after passing the first critical field, related to the configurations that differ in the number of vortices. Overlapping means that, as in the case of the SC-FM prism, we can stabilize a different number of vortices by relaxing the system in a given field. This allows us to assume that the existence of several local energy minima in the same inhomogeneous field, corresponding to different vortex configurations, is a general property.

\section{\label{sec:Conc}CONCLUSIONS}

By solving the TDGL equations, we have numerically studied the nucleation and arrangement of SC vortices in the hybrid structure composed of an SC prism coupled at a distance with an FM prism of the same square cross section. The system under study was small in size, on the order of a few correlation lengths.

We found that vortex nucleation in the non-uniform field of the FM prism begins with small curved vortices at the lower edges of the SC prism, just above the FM. The vortices then gradually increase in length towards the top face of the SC prism, creeping along its sides. This creep-like deformation results from the impact of the stray field on the SC prism and vortex-vortex interactions -- this stage of dynamic is noticeably slower compared to the case of the SC prism placed in a uniform field. Eventually, some of the vortices reach the upper surface and form columnar vortices located in the central part of the prism's vertical cross-section. We observed that the vortices follow the lines of the external magnetic field and provide an intuitive explanation of this effect.

We noticed that the SC prism can reach one of a few stationary states differing in the number of columnar vortices when relaxed in a non-uniform external field of a given strength. Such ambiguity in the selection of the stationary state is not observed for the SC prism exposed to a uniform field. We think that for the SC nanodots placed in a non-uniform field, the pinning of the vortices is stronger and the vortex configuration can be more easily locked in one of the few competing energy minima than in the case of the SC system placed in a uniform field. Furthermore, we found that the presence of columnar vortices is essential for the deterioration of the demagnetizing properties of the SC prism. This is manifested by a rather strong reduction of the average magnetization.

This work provides insights into the dynamic and stationary behavior of vortices in SC-FM hybrids, with implications for optimizing superconducting devices in quantum technologies and spintronics. The provided knowledge is also essential in the context of the recent experimental realization of the SC 3D nanostructures,\cite{zhakina2024} where it is shown that the curvature of the vortices is strongly influenced by the orientation and strength of the applied magnetic field, with higher magnetic fields tending to straighten the vortices. Future research could experimentally validate these findings in the SC-FM hybrids and explore enhanced control over vortex dynamics for improved device performance.

\section*{\label{sec:AUTH}AUTHOR INFORMATION}
\subsection*{\label{sec:Contribut}Author Contributions}
J.K. conceived the initial idea for the project and supervised the research. S.M. developed the core concepts for the simulations and performed the computations. M.K. provided additional supervision and guidance throughout the project. M.G. assisted in verifying the simulations and designed the figures. All authors discussed the results and contributed to the final version of the manuscript.
\subsection*{\label{sec:Notes}Notes}
The authors declare no competing financial interest.

\section*{\label{sec:ACK}ACKNOWLEDGMENTS}
The authors would like to thank A.~Gulian and O.~Dobrovolskiy for fruitful discussions. The work was supported by the grants of the National Science Center – Poland, No. UMO-2021/43/I/ST3/00550 (SM and JWK) and UMO-2020/39/I/ST3/02413 (MG and MK). 

\section*{DATA AVAILABILITY}
Data supporting this study are openly available from the repository at \url{https://doi.org/10.5281/zenodo.14045919}.
\bibliography{main}

\end{document}